\documentclass[10pt]{iopart}

\usepackage[utf8]{inputenc}
\usepackage[pdftex]{graphicx,xcolor}
\usepackage{siunitx}

\let\eqref\ref
\let\citet\cite
\renewcommand{\etal}{\,\textit{et al.}~}

\newcommand{\change}[1]{{\color{orange}#1}}

\begin{document}

\title{Point defects stabilise cubic Mo-N and Ta-N}

\author{Nikola Koutná$^{1,2}$, David Holec$^{3}$, Ondřej Svoboda$^{4}$, Fedor F. Klimashin$^{1}$, and Paul H. Mayrhofer$^{1}$}

\address{$^1$ Institute of Materials Science and Technology, TU Wien, Getreidemarkt 9, A-1060 Vienna, Austria}
\address{$^2$ Faculty of Science, Masaryk University, Kotl\'a\v{r}sk\'a 2, CZ-611 37 Brno, Czech Republic}
\address{$^3$ Department of Physical Metallurgy and Materials Testing, Montanuniversität Leoben, Franz-Josef-Straße 18, Leoben A-8700, Austria}
\address{$^4$ Faculty of Mechanical Engineering, Brno University of Technology, Technick\'a 2, CZ-616 69 Brno, Czech Republic}

\begin{abstract}
We employ \textit{ab initio} calculations to investigate energetics of point defects in metastable rocksalt cubic Ta-N and Mo-N. Our results reveal a strong tendency to off-stoichiometry, i.e., defected structures are surprisingly predicted to be more stable than perfect ones with 1:1 metal-to-nitrogen stoichiometry. Despite the similarity of Ta-N and Mo-N systems in exhibiting this unusual behaviour, we also point out their crucial differences. While Ta-N significantly favours metal vacancies, Mo-N exhibits similar energies of formation regardless of the vacancy type ($V_\text{Mo}$, $V_\text{N}$) as long as their concentration is below $\approx15\,\text{at.\%}$. The overall lowest energy of formation were obtained for $\text{Ta}_{0.78}\text{N}$ and $\text{Mo}_{0.91}\text{N}$, which are hence predicted to be the most stable compositions. To account for various experimental condition during synthesis, we further evaluated the phase stability as a function of chemical potential of individual species. The proposed phase diagrams reveal four stable compositions, $\text{Mo}_{0.84}\text{N}$, $\text{Mo}_{0.91}\text{N}$, $\text{MoN}_{0.69}$ and $\text{MoN}_{0.44}$, in the case of Mo-N and nine stable compositions in the case of Ta-N indicating the crucial role of metal under-stoichiometry, since $\text{Ta}_{0.75}\text{N}$ and $\text{Ta}_{0.78}\text{N}$ significantly dominate the diagram. This is particularly important for understanding and designing experiments using non-equilibrium deposition techniques. Finally, we discuss a role of defects ordering and estimate a cubic lattice parameter as a function of a defect contents and put them in a context of existing literature theoretical and experimental data. 
\end{abstract}


\vspace{2pc}

\submitto{\JPD}

\ioptwocol

\section{Introduction}
Point defects are unavoidably present on materials. At finite temperatures, their concentration is given by thermodynamical equilibrium. However, their amount can be significantly larger, e.g., due to non-equilibrium deposition techniques such as physical vapour deposition (PVD) typically used for synthesis of nitrides. Since they can significantly affect phase stability, composition, and/or material properties, their detailed understanding and precise control become crucial for interpreting experimental data and designing materials with specific properties.

Transition metal nitrides represent a technologically important series of materials and have attracted considerable attention due to their unique mechanical, electrical and chemical properties \cite{Cao}. TaN thin films are desirable as diffusion barrier coating, resistance film, line material in large-scale integrated circuits (LSI), and used in electronic industry \cite{Hashizume}. Compared to the other stable phases, namely the he\-xa\-go\-nal $\pi$-TaN (space group \change{$P\bar{6}2m$}, No:189) and $\theta$-TaN (space group of \change{$P\bar{6}m2$}, No:187), the cubic TaN---appearing at high temperatures and low pressures---exhibits the highest hardness of 30-32\;GPa in the group of transition metal mononitrides \cite{Kieffer}, high bulk modulus \cite{Chang} and superconductivity \cite{Liu, Grumski}. Similarly, the unique combination of physical and chemical characteristics stemming from the wide range of MoN stoichiometries\change{, as investigated recently by Yu\etal\citet{Yu2016-yv},} makes this material very promising for various applications \cite{Jauberteau}, especially as a wear protective coating and diffusion barrier for Al metallizations in ultra large-scale integrated circuits (ULSI) \cite{Hones}. Besides, Jauberteau\etal\citet{Jauberteau} predicted that the cubic MoN can be a potential candidate for a high temperature superconductor with the highest superconducting temperature (30\;K \cite{Papaconstantopoulos}) among all refractory carbides and nitrides.

\change{Both TaN and MoN in their stoichiometric variants prefer hexagonal structures: $\delta$-MoN ($P6_3/mmc$, No:194 \cite{Jauberteau}) and $\pi$-TaN. It is also interesting to note, however, the significant difference between these two structures. All N site and all Mo sites are equivalent in the $\delta$-MoN, each of them having 6 nearest neighbours of the opposite type. Similarly, all N sites in $\pi$-TaN have equivalent neighbourhoods of 5 Ta atoms, i.e., 1 bond less than in the case of MoN. Moreover, there are two types of Ta sites: $1/3$ of them is 3-coordinated and $2/3$ of them being 6-coordinated. These structural differences, stemming from different valence configurations of both metals, are expected to yield also different behaviour of other polymorphs of MoN and TaN.}

Previous theoretical and experimental studies concluded that the NaCl-type MoN and TaN were metastable structures \cite{Grumski, Kanoun, Stampfl,Hones, Jauberteau, Levy, Stampfl2005}, moreover with a wide range of compositions around the 1:1 metal-to-nitrogen stoichiometry. 
Namely, the rocksalt TaN has been predicted to be stable for N/Ta atomic ratio ranging from 0.94 to 1.37 \cite{Shin, Mistrik}. 


As already proposed, cubic Mo-N and Ta-N systems can be stabilized by the vacancies introduction \cite{Grumski, Stampfl, Jauberteau, Hones, Stampfl2005, Violet}, although a detailed \change{theoretical} analysis of the impact of point defects on the phase stability in these two systems is still missing. Furthermore, Jauberteau\etal\citet{Jauberteau} noted that experimentally 
the stoichiometry of MoN compounds depends very sensitively on the nitrogen partial pressure used during the reactive sputtering deposited. Similar observations have been reported also for TaN \cite{Kaul, Yu}. Therefore, the aim of present study is to carefully consider a wide spectrum of defected structures, involving various point defect configurations, obtain the energetics with respect to the occupancy of the metal and N sublattice, and estimated the ranges of stability for various metastable cubic phases \change{which are of relevance for experiment}.

\section{Calculation details}
The Vienna Ab-initio Simulation package (VASP) \cite{VASP_code1, VASP_code2} was used to perform the DFT calculations, employing the projector augmented plane wave (PAW) pseudopotentials under the generalized gradient approximation (GGA) \cite{Kohn} with a Perdew-Burke-Ernzerhof (PBE) exchange correlation functional \cite{PhysRevLett.77.3865}. \change{In particular, their valence configuration (in accordance with the recommendation of the VASP User's Guide) was $4p^65s^24d^4$ for Mo, $5p^65d^36s^2$ for Ta, $2s^22p^3$ for N}. We investigated cubic $\delta$-MoN and $\gamma\text{-}\mathrm{TaN}$ (NaCl structures with space group of $Fm\bar{3}m$, no. 225) which are hereafter referred to as c-MoN and c-TaN. The bulk constants were optimized by fitting the energy vs. volume data with the Birch-Murnaghan equation of state \cite{PhysRev.71.809}. The plane-wave cut\-off energy and the $k$-vector sampling of the Brillouin zone, listed in Tab.~\ref{Details_of_calculation}, were checked to provide a total energy accuracy of about $10^{-3}\;\mathrm{eV/at.}$. A \change{2$\times$2$\times$2} supercell consisting of 64 lattice sites was chosen to model the defected systems. Structure optimizations in defected supercells were carried out by relaxing supercell volumes, shapes, and atomic positions.  

\begin{table}[h!t!]
\centering
\begin{tabular}{ccc}
\hline
\hline
Structure & $E_{\mathrm{cut}}\;[\mathrm{eV}]$ & $k$-point sampling \tabularnewline
\hline
c-TaN & 500 & 6$\times$6$\times$6 \tabularnewline
c-MoN & 700 & 8$\times$8$\times$8 \tabularnewline
\hline
\hline
\end{tabular}
\caption{Overview of plane-wave cutoff energies and the Monkhorst-Pack \cite{PhysRevB.13.5188} sampling of Brilouin zone used in this work.}
\label{Details_of_calculation}
\end{table}

Various defect configurations were investigated, ranging from fully disordered to fully ordered configurations. The former ones were generated according to the special qua\-si\-ran\-dom structure (SQS) method \cite{SQS}, 
while the later ones were constructed from a conventional cubic B1 cell (8 lattice sites) containing one or two N vacancies, and subsequently expanded to $2\times 2\times 2$ supercells. The same approach as used for fully ordered structures here has been previously proposed by Grumski\etal\citet{Grumski}. To obtain various partially ordered structures, we started from the above mentioned defected ordered supercells, and arbitrarily filled as many vacancy sites as needed to obtain the desired composition. Finally, the degree of order of various structures was quantified with the Warren-Cowley short-range order (SRO) parameter \cite{Cowley}. Denoting $j$ the coordination shell, the corresponding SRO parameter $\alpha^j$ is defined as
\begin{equation}
\alpha^j=1-\frac{N^j_{AB}}{Nx_AM^jx_{AB}},
\label{SRO_parameter}
\end{equation}
where $N$ is the number of atoms in supercell, $N^j_{AB}$ is the number of bonds between atoms A and atoms B corresponding to the coordination shell $j$, $x_A$ and $x_B$ stands for concentration of atoms A and B, respectively, and $M^j$ is the coordination number of the shell $j$. In the present study, we denote 
\begin{equation}
\alpha:=\sum_{j=1}^n|\alpha^j| \label{eq:alpha}
\end{equation}
the sum of the absolute values of SRO parameters and use it to measure the degree of structural order when comparing different configurations with the same vacancy content. 

The analysis of relative chemical stability is based on discussing the energy of formation, $E_f$, calculated as
\begin{equation}
E_f=\frac{E_{\mathrm{tot}}-\sum_{s}n_s\mu_s}{\sum_sn_s},\label{eq:Ef}
\end{equation}
where $E_{\mathrm{tot}}$ is the total energy of (defected) supercell, $n_s$ and $\mu_s$  are the number of atoms and the chemical potential, respectively, of atom specie $s=$Mo/Ta or N. The reference chemical potentials for Mo, Ta, and N are conventionally set to that of bcc-Mo, $\mu_{\text{Mo}}(\text{bcc-Mo)}$, bcc-Ta, $\mu_{\text{Ta}}(\text{bcc-Ta)}$, and N$_2$ molecule, $\mu_{\text{N}}(\text{N}_2)$, respectively.

Depending on the specific experimental conditions, their values may however differ.
Thus in the later part of the manuscript, we evaluate the energy of formation as a function of $\mu_{\text{N}}$ and $\mu_{\text{Mo}}$, or $\mu_{\text{N}}$ and $\mu_{\text{Ta}}$, considering their upper limits
\begin{eqnarray}
\mu_{\mathrm{N}} - \mu_{\mathrm{N}}(\mathrm{N}_2) \leq 0\ , \label{eq:mu_N limit} \\
\mu_{\mathrm{Mo}} - \mu_{\mathrm{Mo}}(\text{bcc-Mo}) \leq 0\ , \label{eq:mu_Mo limit}\\
\mu_{\mathrm{Ta}} - \mu_{\mathrm{Ta}}(\text{bcc-Ta}) \leq 0\ . \label{eq:mu_Ta limit}
\end{eqnarray}
which are hereafter termed as the N-rich, Mo-rich, and Ta-rich conditions, respectively. The Ta-rich (Mo-rich)/N-poor state will be approached when there is excess of Ta (Mo) atoms available in the system as compared with the N molecules, while the opposite  holds for Ta-poor (Mo-poor)/N-rich conditions \cite{Stampfl2005}. 
A structure is predicted to be unstable for a combination of $\mu_{\text{N}}$ and $\mu_{\text{Mo}}$, or $\mu_{\text{N}}$ and $\mu_{\text{Ta}}$ yielding a positive value of $E_f$ (cf.~Eq.\eqref{eq:Ef}). \change{In order to provide a more specific meaning to an otherwise abstract value of $\mu_{\text{N}}$, we adopted the approach form Refs.~\cite{Reuter2001-fj,Stampfl}, which expresses temperature and pressure dependence of the chemical potential as
\begin{eqnarray}
  \mu_{\mathrm{N}}(T,p_{\mathrm{N}_2})=&\frac12\left(\Delta\mu(T,p_{{\mathrm N}_2}^0)+k_BT\ln\frac{p_{\mathrm{N}_2}}{p_{\mathrm{N}_2}^0}+\right.\nonumber\\
  &\left.\phantom{\frac12}+E_{\mathrm{tot}}(\mathrm{N}_2)\right)\ .\label{eq:mu}
\end{eqnarray}
$E_{\mathrm{tot}}(\mathrm{N}_2)$ in the above expression is the total energy of N$_2$ molecule at 0\,K, $p_{{\mathrm N}_2}^0$ is a reference pressure (typically 1\,atm) and $p_{{\mathrm N}_2}$ is the actual pressure of N$_2$ gas.\begin{eqnarray}
  \Delta\mu(T,p_{{\mathrm N}_2}^0)=& H(T,p_{{\mathrm N}_2}^0)-H(0\,\mbox{K},p_{{\mathrm N}_2}^0)-\nonumber\\
  &-T\left(S(T,p_{{\mathrm N}_2}^0)-S(0\,\mbox{K},p_{{\mathrm N}_2}^0)\right)\ ,\label{eq:deltamu}
\end{eqnarray}
where $H$ and $S$ are tabulated values of Helmholtz free energy and entropy of N$_2$ at temperatures $T$ or 0\,K, and reference pressure $p_{{\mathrm N}_2}^0$, taken from Ref.~\cite{Stull1971-tu}.
}

Furthermore, to quantify the energy cost of vacancy introduction in a perfect material, vacancy formation energy, $E_{f,\mathrm{ vac}}$ was calculated according to
\begin{equation}
E_{f,\mathrm{vac}}=E_{\mathrm{tot}}^{\mathrm{def}}-E_{\mathrm{tot}}^{\mathrm{perf}}+\sum_s n^{\mathrm{vac}}_s\mu_s,\label{eq:Efvac}
\end{equation}
where $E_{\mathrm{tot}}^{\mathrm{perf}}$ and $E_{\mathrm{tot}}^{\mathrm{def}}$ are the total energies corresponding to the perfect and defected supercell, respectively, and $n^{\mathrm{vac}}_s$ denotes the number of missing atomic species $s$.

\section{Results and discussion}

\subsection{Stability of cubic Mo-N and Ta-N systems}
\begin{figure}[h!]
	\centering
    \includegraphics[width=8.5cm]{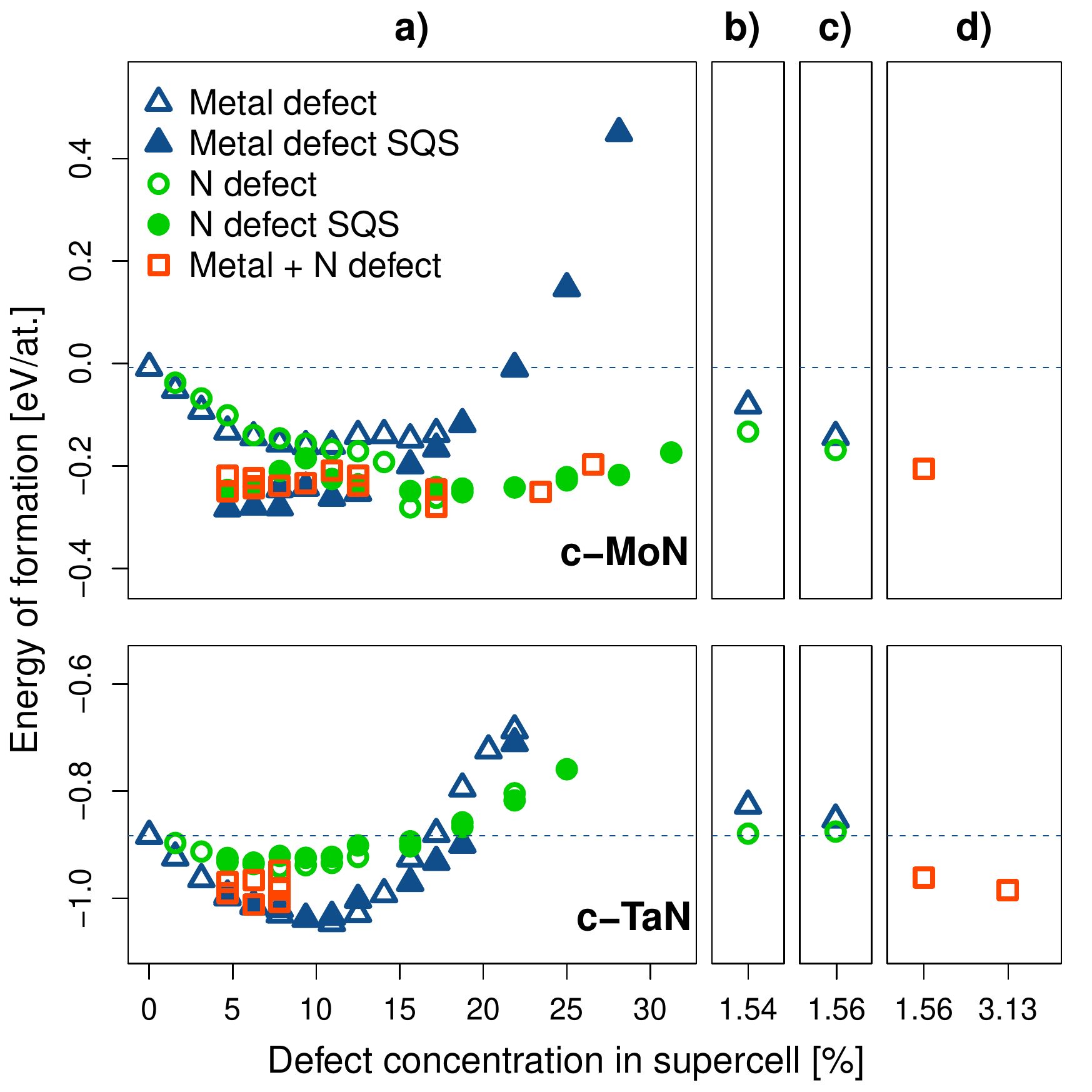}
	\caption{Energy of formation, $E_f$, as a function of the defect concentration in c-MoN (upper panels) and c-TaN (lower panels). Defects are sorted according to their type:  a) vacancies, b) interstitials, c) Frenkel pairs, and d) Schottky defects. The two horizontal lines guide the eye for $E_f$ of perfect c-MoN and c-TaN, respectively.}
\label{FIGURE_01}
\end{figure}

Energy of formation as a function of the defect concentration (see Fig.~\ref{FIGURE_01}) suggests that in the case of c-TaN, Ta vacancies are preferred over N vacancies up to the concentration $\sim 16\%$ ($\text{Ta}_{0.69}\text{N}$), which agrees well with the previously published predictions of Grumski\etal\citet{Grumski}, and Stampfl and Freeman \citet{Stampfl}. Interestingly, also N vacancies up to the concentration $\sim 13\%$ ($\mathrm{TaN}_{0.75}$) are found slightly more energetically favourable then the perfect TaN system. As intuitively expected, $E_f$ of structures containing mixed Ta and N vacancies lie between $E_f$ of structures containing only Ta or only N vacancies.

Consistently with previous studies \cite{Stampfl, Zhao}, $\mathrm{Ta}_{0.78}\mathrm{N}$ ($\sim 11\%$ Ta vacancies) is the energetically most preferred structure; its formation energy $-1.05\;\mathrm{eV}/\mathrm{at.}$ lies about $0.17\;\mathrm{eV}/\mathrm{at.}$ lower than $E_f$ of the perfect c-TaN. When restricted to only N vacancies, the energy of formation reaches its minimum for $\mathrm{TaN}_{0.84}$ ($\sim 8\%$ N vacancies), however, this value is only $0.02\,\mathrm{eV/at.}$ lower than $E_f$ of the perfect structure. 

The energy of formation can be considerably decreased even when keeping the 1:1 metal-to-nitrogen stoichiometry, by introducing Schottky defects. Such a configuration is practically undetectable by conventional tools used for chemical analysis, e.g., energy-dispersive X-ray spectroscopy (EDX) or Rutherford backscattering spectroscopy (RBS). As discussed in the next section, structures containing Schottky defects yield smaller lattice constant (as estimated by, e.g. X-ray diffraction (XRD)) and hence may contribute to relatively large scatter of experimental data (cf. Tab.~\ref{Lattice_parameters}). Finally, our calculations suggest that both interstitial and Frenkel pairs destabilize c-TaN, which is in agreement with previous study of Stampfl and Freeman~\citet{Stampfl}. In\-ter\-est\-ing\-ly, the formation energy of Frenkel pairs was found to be dependent on the distance between the missing and the interstitial atom sites. Assuming this distance generally smaller than $2\si\angstrom$-$3\si\angstrom$, we speculate that the interstitial atom attempts to fill the vacancy site, thus the structure finally ends up in its almost perfect configuration. However, considering larger distance between vacancy and interstitial, the energy of formation remains almost constant. Hence only data points for the later case are plotted in Fig.~\ref{FIGURE_01} (analogically in the case of c-MoN).

Concerning c-MoN, similar but more complex behaviour is predicted. Formation energy cannot be simply categorised according to the vacancy type, but instead seems to be governed by the ordering degree of the defects. Details will be discussed in Section~\ref{sec:ordering}, here we focus on overall compositional trends. The $E_f$ global minimum ($-0.282\;\mathrm{eV/at.}$) is obtained for $\sim 4.69\%$ fully disordered metal vacancies ($\mathrm{Mo}_{0.91}\mathrm{N}$), nevertheless, a partially ordered structure with $\sim 15.63\%$ of N vacancies ($\mathrm{MoN}_{0.69}$) exhibits a very close value of $-0.280\;\mathrm{eV/at.}$. Furthermore, in contrast to c-TaN, not only Schottky defects, but also interstitials and Frenkel pairs may decrease the formation energy with respect to the perfect c-MoN. Namely, the energy of formation in the case of a N interstitial ($\mathrm{MoN}_{1.03}$) is comparable with that of $\mathrm{MoN}_{0.84}$ containing $\sim 7\%$ N vacancies. Slightly lower are found Mo and N Frenkel pairs, and the lowest $E_f$ is predicted for Schottky defects yielding $E_f=-0.21\;\mathrm{eV/at.}$ We speculate that this behaviour can be ascribed to the fact that both, metal and N vacancies are strongly favoured over the perfect structure (unlike the c-TaN case where the metal vacancies are significantly preferred over the N ones). Consequently, almost arbitrary combination of point defects at low concentration results in an increased stability of the Mo-N system.

\subsection{Impact of deposition conditions}

\begin{figure*}[t]
	\centering    	
	\includegraphics[width=13cm]{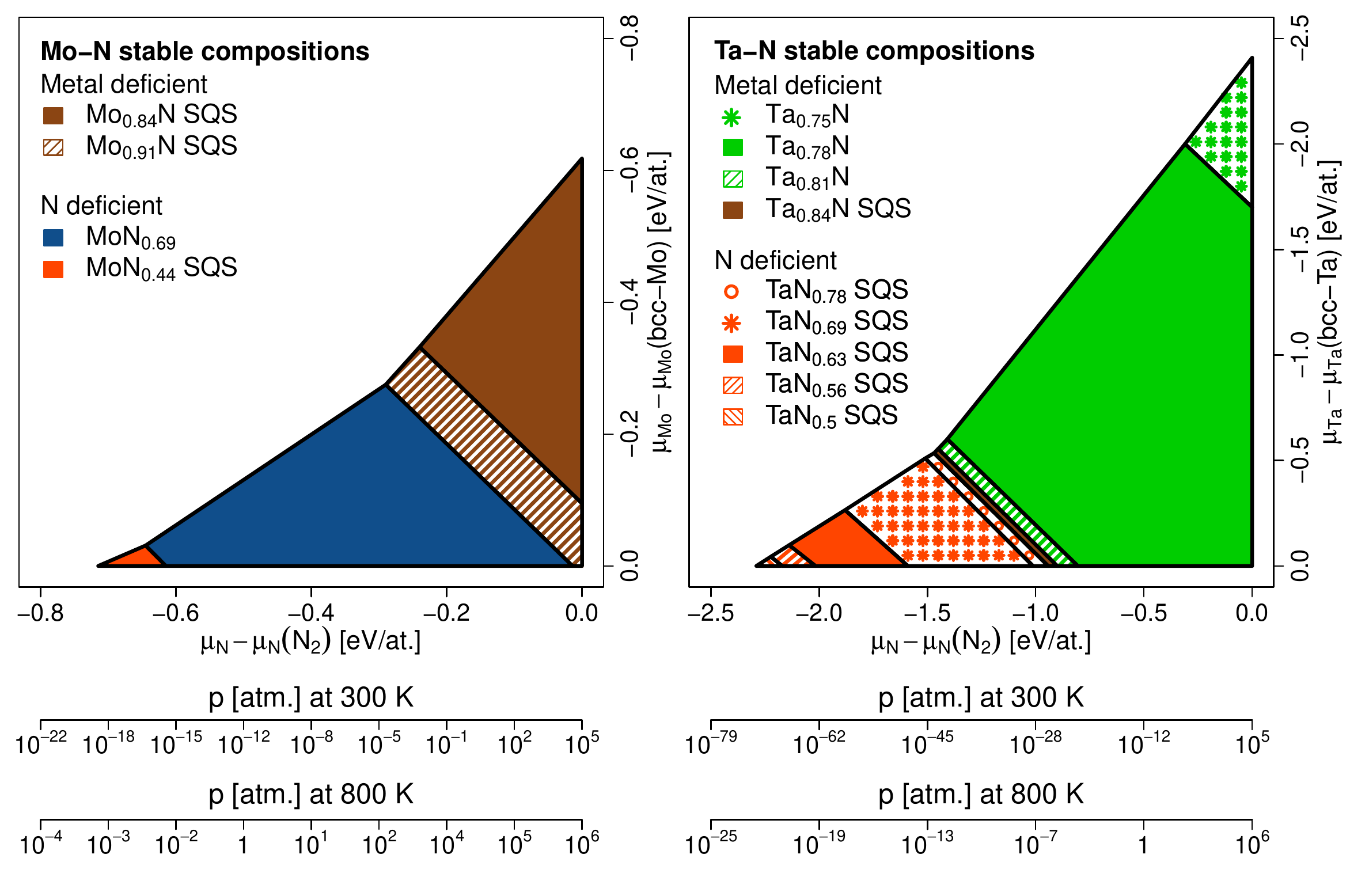}
\caption{The cubic Mo-N (left) and Ta-N (right) phase diagram presenting the stable compositions as a function of the N, Ta and Mo chemical potentials. The uncolored area corresponds to unstale compositions.}
\label{FIGURE_05}
\end{figure*}

Since it has been experimentally reported that the MoN and TaN stoichiometry sensitively depends on the deposition conditions \cite{Jauberteau, Kaul, Yu}, we analyse the role of chemical potentials. The energy of formation $E_f$ is considered as a function of two independent variables, $\mu_{\text{N}}$ and $\mu_{\text{Mo}}$ in the case of c-MoN, $\mu_{\text{N}}$ and $\mu_{\text{Ta}}$ in the case of c-TaN. 
The most stable structures, i.e., the structures with the lowest $E_f$, are depicted in Fig.~\ref{FIGURE_05} as a function of the respective chemical potentials. The unshaded regions correspond to unstable structures, i.e., structures yielding a positive value of $E_f$. \change{In order to provide a guidance for future experimental studies, the N chemical potential has been also recalculated to N$_2$ pressure at 300\,K and 800\,K, using Eqs.~\ref{eq:mu} and \ref{eq:deltamu}.}

The stability regions of metal and N vacancies are almost of equal size in the Mo-N case, suggesting that the actual formation of either metal, or nitrogen deficient structure very sensitively depends on the experimental conditions. Four off-stoichiometric structures $\text{Mo}_{0.84}\text{N}$, $\text{Mo}_{0.91}\text{N}$, $\text{MoN}_{0.69}$ and $\text{MoN}_{0.44}$ are predicted to be stable. Interestingly, keeping the Mo-rich conditions while arbitrarily varying $\mu_{\text{N}}$ results in stabilizing only N deficient structures and vice versa, i.e., solely Mo deficient compounds are predicted in the case of N-rich conditions and arbitrary values of $\mu_{\text{Mo}}$. Furthermore, while nitrogen vacancies tend to order, the opposite trend is shown for their metal counterparts.

In contrast, metal deficient structures, namely $\text{Ta}_{0.75}\text{N}$, $\text{Ta}_{0.78}\text{N}$, $\text{Ta}_{0.81}\text{N}$ and $\text{Ta}_{0.84}\text{N}$, dominate the Ta-N phase diagram showing that the (partially) ordered metal vacancies are strongly favoured. We note that $\text{Ta}_{0.75}\text{N}$ represents a cell with fully ordered vacancies (1 Ta missing in every conventional cubic 8-atom cell), and hence is the most ordered among all defected supercells in this study (cf. Eq.~\ref{eq:alpha}). Underpinning our previous results, $\text{Ta}_{0.78}\text{N}$ is predicted to be stable in a remarkably wide range of $\mu_{\text{N}}$ and $\mu_{\text{Ta}}$ values. Our calculations further reveal an array of disordered highly nitrogen deficient compositions: $\text{TaN}_{0.78}$, $\text{TaN}_{0.69}$ and $\text{TaN}_{0.63}$, followed by $\text{TaN}_{0.56}$ and $\text{TaN}_{0.5}$, while the atomic ratio N/Ta decreases as approaching the strongly N-poor conditions.   

Importantly, neither of the two systems Mo-N and Ta-N exhibits a stability region for perfectly stoichiometric metal-to-nitrogen cubic structures. This is in excellent agreement with previous studies showing that the stoichiometric structures are hexagonal $\delta$-MoN (space group of $P6_3mc$, no. 186) \cite{Jauberteau} and $\pi$-TaN (CoSn type structure with space group of $P\bar62m$, no. 189) \cite{Grumski, E_J_Zhao}.

\subsection{Vacancy formation energies}
  
\begin{figure}[h!]
	\centering
    \includegraphics[width=7cm]{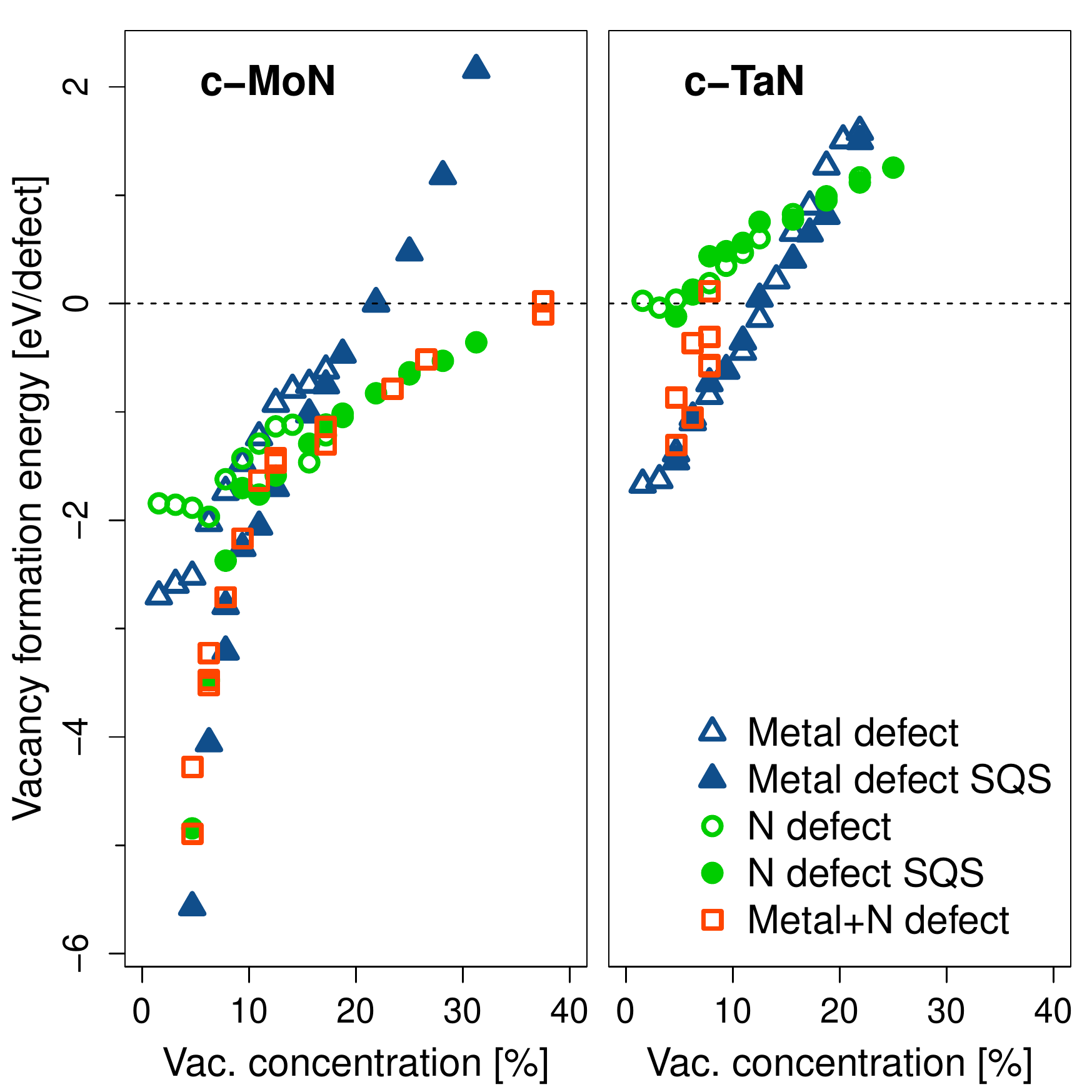}
	\caption{Vacancy formation energy, $E_{f,\mathrm{vac}}$, as a function of vacancy concentration in c-MoN (left panel) and c-TaN (right panel).}
\label{FIGURE_02}
\end{figure}

The negative value of vacancy formation energy, $E_{f,\mathrm{vac}}$, eva\-lu\-ated according to Eq.~\eqref{eq:Efvac}, means that energy is released when forming a vacancy, therefore, that the configuration with the vacancy is energetically preferred to a perfect structure. The trends shown in Fig.~\ref{FIGURE_02} are a different evaluation of the calculations already discussed in the previous sections. It is important to realise, however, that $E_f$ and $E_{f,\mathrm{vac}}$ represent different phenomena. $E_f$ describes chemical stability with variable number of atoms in the supercell (e.g., evaluation of $E_f$ of a structure with $(N-1)$ atoms and 1 vacancy takes into consideration chemical potential of only $(N-1)$ species, cf. Eq.~\eqref{eq:Ef}), hence refer to the formation of perfect or defected state from a reservoir of particles. On the other hand, vacancy formation energy, $E_{f,\mathrm{vac}}$, considers always $N$ species (cf. Eq.~\eqref{eq:Efvac}), and hence describes formation of the defected stated from the perfect structure. 


Both c-MoN and c-TaN exhibit a remarkably broad range of compositions with negative $E_{f,\mathrm{vac}}$, \change{confirming} that off-stoichiometric compounds are strongly favoured at 0\;K \change{(cf.~Fig.~\ref{FIGURE_01})}. The c-TaN phase yields mostly positive (though relatively small) formation energies for N vacancies which is in line with the computational results reported by Grumski\etal\citet{Grumski}. In contrast to c-TaN, values of $E_{f,\mathrm{vac}}$ of c-MoN lie in a surprisingly wide range (here from $-6\;\mathrm{eV/defect}$ to $2\;\mathrm{eV/defect}$). Additionally, comparison of metal and N (partially) ordered and disordered configurations reveals substantial differences in the case of c-MoN, while the degree of order seems not to play any significant role for c-TaN. These predictions underline the qualitatively different incorporation of vacancies in both c-MoN and c-TaN.

\subsection{Structural parameters}

Cubic lattice parameter, $a$ was evaluated as 
\begin{equation}
  a = \sqrt[3]{V_0}\ ,\label{eq:aLat}
\end{equation}
where $V_0$ is the equilibrium volume corresponding to a conventional cubic cell. Our structural models are based on $2\times2\times2$ supercells constructed from the conventional cubic cells, therefore we take $V_0=V/8$ where $V$ is volume of a fully-relaxed supercell with or without defect(s). It is important to note that due to supercell-size effect, the relaxed defected supercells do not have, in general, cubic shape anymore. However, this is a consequence of atomistic approach using only a few dozens of atoms; the defected materials will macroscopically still possess cubic symmetry. Evaluation using Eq.~\eqref{eq:aLat} hence effectively averages these local distortions and restores the macroscopically cubic material.

\begin{figure}[h!]
	\centering
    \includegraphics[width=7cm]{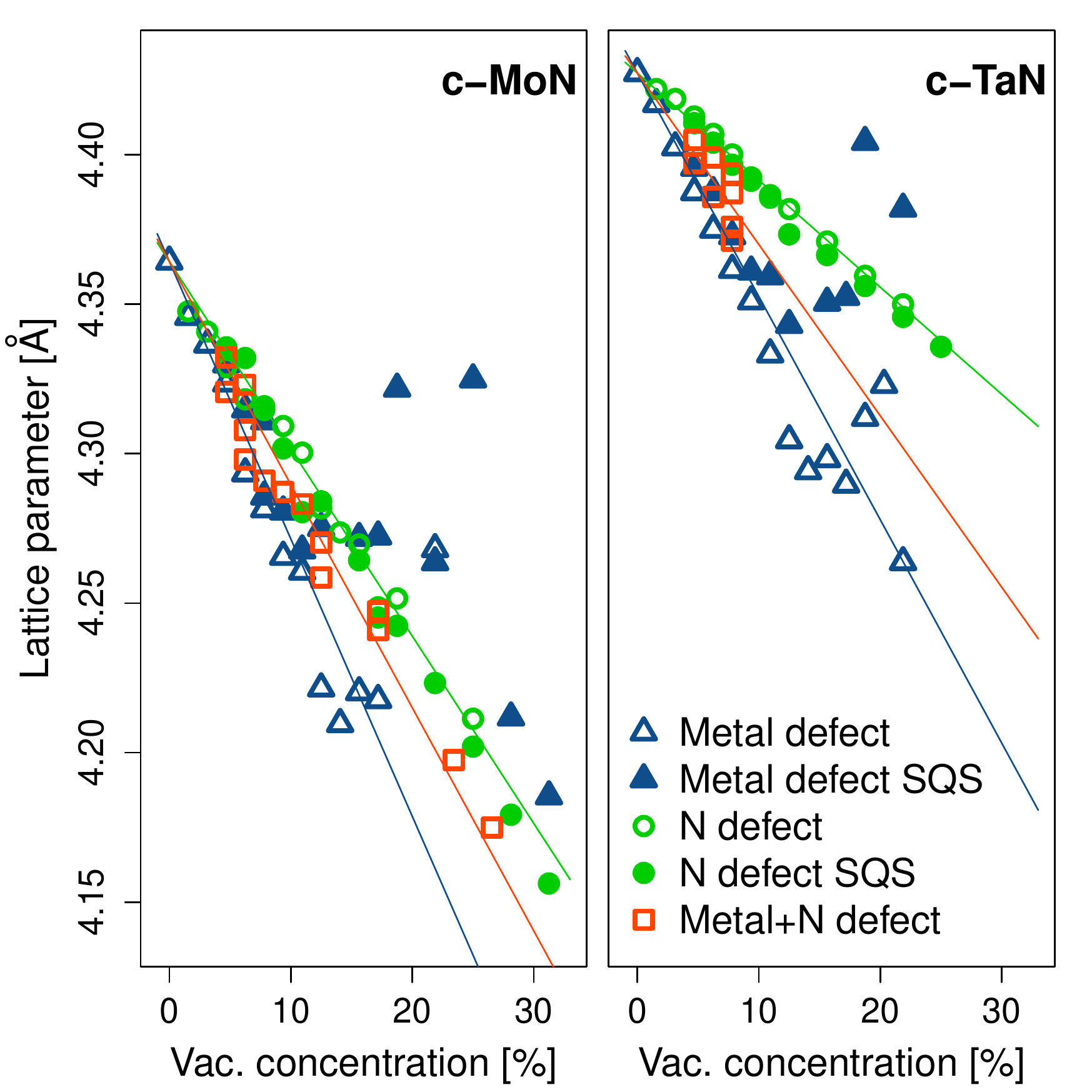}
	\caption{Lattice parameter dependence on concentration of vacancies in c-MoN and c-TaN supercell, respectively. \change{The lines are linear fits according to Eq.~\ref{eq:afit}.}}
\label{FIGURE_03}
\end{figure}

Our calculations for c-MoN and c-TaN illustrate strong decrease of $a$ as the vacancy concentration increases (see Fig.~\ref{FIGURE_03}). Comparable trends were predicted using the first principles for c-MoN by Lowther~\citet{Lowther} and for c-TaN by Stampfl and Freeman~\citet{Stampfl}, and Grumski\etal\citet{Grumski}. Table \ref{Lattice_parameters} lists the calculated lattice parameters for several selected compositions for which experimental or {\it{ab initio}} values were previously published in literature, hence corroborating our predictions.

\change{To provide a quantitative estimation of the lattice parameters, we fitted the calculated values for defect concentrations below $x<0.12$, with linear functions of the defect concentration, $x$, (see Fig.~\ref{FIGURE_03}):
\begin{equation}
  a(x)=a(0)-\beta x\ .\label{eq:afit}
\end{equation}
Here, $a(0)$ is the lattice parameter of perfect (defect-free) TaN or MoN, respectively. The fitted $\beta$ values for MoN are $0.00928$, $0.00627$, and $0.00746\,\mbox{\AA}$ for Mo vacancies, N vacancies, and Schottky defects (Mo+N vacancies), respectively. In the TaN case, $\beta$ takes value of $0.00747$, $0.00358$, and $0.00573\,\mbox{\AA}$ for Ta vacancies, N vacancies, and Schottky defects (Ta+N vacancies), respectively. In general, metal deficient structures show much greater lattice parameter (and hence volume)} decrease upon vacancy introduction than their nitrogen deficient counterparts. 
It is a well-established fact that the GGA and the local density approximation (LDA) exchange-correlation functionals tend to over- and underestimate, respectively, the predicted lattice constants with respect to experimental values. This has been explicitly shown by Grumski\etal\citet{Grumski} to be valid in the case of c-TaN. Consistently with this conclusion, our predictions listed in Tab.~\ref{Lattice_parameters} are slightly higher than the experimental results and the values calculated using the LDA.

\begin{table*}[t]
\caption{Comparison of lattice parameters (in $\si\angstrom$) calculated in this study with corresponding experimental or theoretical values form literature for selected compositions.}
\label{Lattice_parameters}
\centering
\begin{tabular}{ccccccc}
\hline
\hline
&TaN & $\mathrm{Ta}_{0.97}\mathrm{N}$& $\mathrm{Ta}_{0.94}\mathrm{N}$ & $\mathrm{Ta}_{0.87}\mathrm{N}$ & $\mathrm{Ta}_{0.75}\mathrm{N}$ & $\mathrm{Ta}\mathrm{N}_{0.97}$  \tabularnewline
\hline
Present study & 4.427 & 4.417& 4.403 &  4.38  & 4.305 & 4.422  \tabularnewline
Experiment & 4.42 \cite{JCPDS_TaN}, 4.427 \cite{Nie}, 4.336 \cite{Mashimo} & & & 4.361 \cite{JCPDS_TaN_1.13} &  & \tabularnewline
LDA & 4.353 \cite{Grumski}, 4.386 \cite{Chang}, 4.397 \cite{Papaconstantopoulos} & & & & & \tabularnewline
GGA & 4.424 \cite{Grumski}, 4.415 \cite{E_J_Zhao}, 4.414 \cite{Ren},& 4.41 \cite{Grumski}& 4.40 \cite{Grumski} & & 4.30 \cite{Grumski} & 4.42 \cite{Grumski} \tabularnewline
 & 4.547 \cite{Li}, 4.40 \cite{Cao}, 4.408 \cite{Isaev} & & & & &\tabularnewline
\hline
\hline
&MoN & $\mathrm{MoN}_{0.75}$ &  $\mathrm{MoN}_{0.5}$   & & &\tabularnewline
\hline
Present study  & 4.364  &  4.282  & 4.211& \tabularnewline
Experiment & 4.215-4.253 \cite{Savvides}, 4.20-4.22 \cite{Saito}, &   & & & &\tabularnewline
 &  4.20-4.27 \cite{Ihara}  &   & & \tabularnewline
LDA & 4.280 \cite{Lowther}, 4.41 \cite{Grossman}, 4.285 \cite{Chen},  & 4.217 \cite{Lowther}  &    4.162 \cite{Lowther} & & &\tabularnewline
 & 4.25 \cite{Papaconstantopoulos} &   &     & & &\tabularnewline
GGA & 4.328 \cite{Isaev}  &  &    & & &\tabularnewline
\hline
\hline
\end{tabular}
\end{table*}

Finally, lattice parameters corresponding to the other types of point defects investigated in present study were calculated. As a consequence of introducing one interstitial atom into c-MoN or c-TaN, lattice expands to accommodate the additional atom. Interestingly, similar lattice expansion is predicted also for Frenkel pairs. On the other hand, Schottky defects lead to a decrease of the lattice constant. This can be intuitively understood by the fact that Schottky defects are complexes of metal and N vacancies, which both lead to a lattice parameter decrease (cf.~Fig.~\ref{FIGURE_03}).

\subsection{Ordering of vacancies}\label{sec:ordering}

\begin{figure}[h!]
	\centering    	
	\includegraphics[width=8cm]{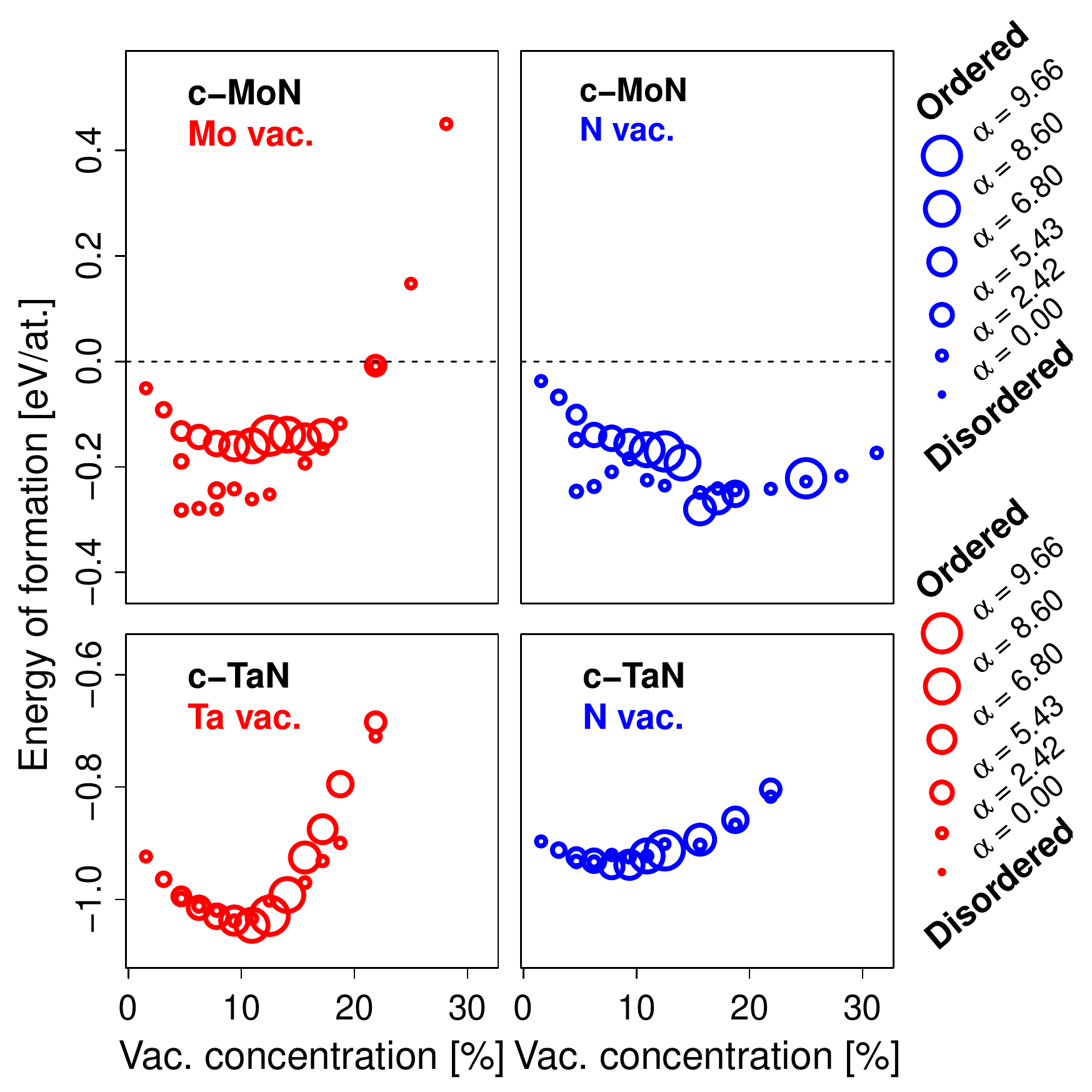}
\caption{Formation energy of metal (left panels) and N (right panels) vacancies in c-MoN (upper panels) and c-TaN (lower panels), as a function of vacancy concentration and the degree of order quantified by $\alpha := \sum_{j=1}^n|\alpha^j|$, where $\alpha^j$ are the SRO parameters defined by Eq.~\ref{SRO_parameter}.}
\label{FIGURE_04}
\end{figure}

A spectrum of structures ranging from fully disordered to (partially) ordered has been considered in this work, which allows to discuss the results with respect to the structural degree of order. We note that the data presented in Fig.~\ref{FIGURE_04} describe the case of N-rich and Mo-rich or N-rich and Ta-rich conditions, respectively. Our analysis clearly demonstrates that the key difference between $\text{c-MoN}$ and $\text{c-TaN}$ lies in their opposite preference for ordering/disordering vacancies. While the metal vacancies in c-MoN fa\-vour rather disordered configurations (differing from ordered and partially ordered by $0\text{--}0.15\;\mathrm{eV/at.}$), the partially ordered metal vacancies in c-TaN exhibit lower values of formation energy up to the concentration $\sim 14\%$ ($\text{Ta}_{0.72}\text{N}$). Concerning N vacancies in $\text{c-TaN}$, the same trend as for their metal counterparts is observed. 
Finally, an interesting behaviour is predicted for N vacancies in c-MoN. Generally, disordered configuration is preferred, although an ordered arrangement is favoured between $16\%\text{--}22\%$ of N vacancies ($\text{MoN}_{0.69}\text{--}\text{MoN}_{0.44}$). We also conclude that the impact of the vacancy degree of order on the stability significantly decreases with increasing non-stoichiometry in both cases, as demonstrated by the energy of formation being almost independent of the order parameter (Eq.~\ref{eq:alpha}).

\section{Conclusions}
We have carried out detailed first-principles investigations on the energetics of point defects in cubic B1-type MoN and TaN. The results demonstrate a strong tendency for off-stoichiometry in both systems which is in excellent agreement with previous studies \change{on TaN,} and experimental observations reporting $\text{c-MoN}$ and $\text{c-TaN}$ with 1:1 metal-to-nitrogen stoichiometry as metastable structures. Our calculations indicate that phase stability can be largely affected by the vacancies formation. Although the fact, that defected structures are more stable than their perfect parent phases, is fascinating on its own, we have additionally \change{newly} revealed differences in ordering tendencies of point defects in both, Mo-N and Ta-N systems.

Namely, metal vacancies in concentration $\sim 11\%$ result in the most stable composition Ta$_{0.78}$N, while $\sim 5\%$ of metal vacancies or $\sim\;16\%$ of N vacancies, i.e. Mo$_{0.91}$N or MoN$_{0.69}$, is found the most favourable for Mo-N system.  We proposed phase diagrams revealing a spectrum of stable compositions depending on the actual values of chemical potentials. The Ta-N phase diagram is dominated by metal deficient structures, $\text{Ta}_{0.75}\text{N}$, $\text{Ta}_{0.78}\text{N}$, $\text{Ta}_{0.81}\text{N}$, $\text{Ta}_{0.84}\text{N}$, especially indicating that the largest stability region corresponds to (partially) ordered $\text{Ta}_{0.78}\text{N}$. When approaching the N-poor conditions (while staying close to the Ta-rich conditions), an array of N deficient compounds is predicted, with N content rapidly decreasing (from  $78\%$ to $50\%$ concentration in the sublattice) as approaching the strong N-poor conditions. In the Mo-N case, only four stable compositions are predicted, $\text{Mo}_{0.84}\text{N}$, $\text{Mo}_{0.91}\text{N}$, $\text{MoN}_{0.69}$ and $\text{MoN}_{0.44}$. The stability regions are shown to be evenly distributed between N over- and under-stoichiometric structures. 
The predicted metastable configurations of metal-deficient MoN and TaN are predicted to tend to disorder and order, respectively.

Finally, strong dependence of the cubic lattice parameter on the actual composition was clearly \change{demonstrated and a compositional dependence of the cubic lattice parameter on the vacancy content was provided}. Defected structures with various degrees of order were compared in terms of their formation energy. This together with the negative energy of formation of Schottky defects in TaN is suggested to be responsible for the large scatter of reported experimental lattice parameters.

\section*{Acknowledgements}
Access to computing and storage facilities owned by parties and projects contributing to the National Grid Infrastructure MetaCentrum, provided under the programme "Projects of Large Infrastructure for Research, Development, and Innovations" (LM2010005), is greatly appreciated. The computational results presented have been achieved in part using the Vienna Scientific Cluster (VSC). Financial support by the START Program (Y371) of the Austrian Science Fund (FWF) is also greatly acknowledged.

\section*{References}

\providecommand{\newblock}{}

\end{document}